# Proposing Cluster_Similarity Method in Order to Find as Much Better Similarities in Databases

Mohammad-Reza Feizi-Derakhshi[1] and Azade Roohany[2]

[1]**Department of Computer, University of Tabriz, Tabriz, Iran**
*mfeizi@tabrizu.ac.ir*

[2] **Department of Computer, Shabestar Branch, Islamic Azad University, Shabestar, Iran**
*roohany_azade@yahoo.com*

**Abstract**

Different ways of entering data into databases result in duplicate records that cause increasing of databases' size. This is a fact that we cannot ignore it easily. There are several methods that are used for this purpose. In this paper, we have tried to increase the accuracy of operations by using cluster similarity instead of direct similarity of fields. So that clustering is done on fields of database and according to accomplished clustering on fields, similarity degree of records is obtained. In this method by using present information in database, more logical similarity is obtained for deficient information that in general, the method of cluster similarity could improve operations 24% compared with previous methods.

*Keywords:* *Clustering, cluster similarity, Record similarity, Field similarity*

## 1. Introduction

Nowadays the size of databases is increasing by developing of information and advancing of technology and requirement for proper and accurate restoration of necessary information has become an important issue in this field. One of the matters that are introduced in most sources is the lack of compatibility of identical data in databases that despite of same meaning, they stored in different shapes that is resulted from improper entering of data such as type errors, the way of speaking, abstraction and etc.

Duplicate records are unfavorable [1]. The main fact is that how we can eliminate similar records. Such fact is called record linkage or record matching. This is the task of accurate labeling of pair records that are related to same entity from different sources [2]. In other words the aim of one record linkage algorithm is to detect records which do not have complete matching, but they have some similarities. By finding similar records, we can combine them that help to decrease the size of databases. In the next sections of present paper, first we will discuss about finding of similar records and previous methods and in the next section we will describe the steps of proposed algorithm and finally we will present conclusion.

## 2. Finding similar records

We can identify and combine similar records by using some methods in order to minimize the size of databases. So first, field matching algorithms take two fields as an input and then they return their similarity in the numerical format between one and zero. After that the detection of similar records among records of database is done according to obtained numbers and finally clustering is done based on obtained similarities of records.

One of the previous algorithms that we need is *Jaro* algorithm that is used for field similarity. One clustering algorithm is also needed in order to use single linkage method that will be described.

2.1 Jaro Distance Metric

Jaro introduced a string comparison algorithm that was mainly used for comparison of last and first names. The basic algorithm for computing the Jaro metric for two strings S1 and S2 includes the following steps:
 1. Compute the string lengths |S1| and |S2|.
 2. Find the "common characters" c in the two strings; common are all the characters S1[j] and S2[j] in (1)

$$|i-j| <= 1/2 \min \{ |S1|, |S2| \} \quad (1)$$

 3. Find the number of transpositions t; the number of transpositions is computed as follows: We compare the ith common character in S1 with the ith common character in S2. Each non matching character is a transposition. The Jaro comparison value is calculated by equation (2).

$$Jaro(|S1|,|S2|) = \frac{1}{3}(\frac{c}{|S1|} + \frac{c}{|S2|} + \frac{c - \frac{t}{2}}{c}) \quad (2)$$

From the description of the Jaro algorithm, we can see that the Jaro algorithm requires O(|S1| *|S2|) time for two strings of length |S1| and |S2|, mainly due to Step 2, which computes the "common characters" in the two strings. Winkler and Thibaudeau modified the Jaro metric to give higher weight to prefix matches since prefix matches are generally more important for surname matching [1,3].

2.2 Single linkage method

Single linkage method is one of the oldest and simplest clustering methods and also is one of the hierarchical and individual clustering methods. We can also call it nearest neighbor and either connectedness method or minimum method. That by assuming that B and A are two clusters, according to figure 1, distance d(A, B) equals to at least the distance between correspondent patterns of B and A that is calculated by equation (3).

$$d(A,B) = \min_{i \in A, j \in B} d(i.j) \quad (3)$$

In this method the nearest distance between two clusters is considered. Clustering based on this distance is one of the most common methods in clustering. Since this algorithm is hierarchical, when clusters are combined in order to form new clusters, it erases correspondent rows and columns in the adjacency matrix [4].

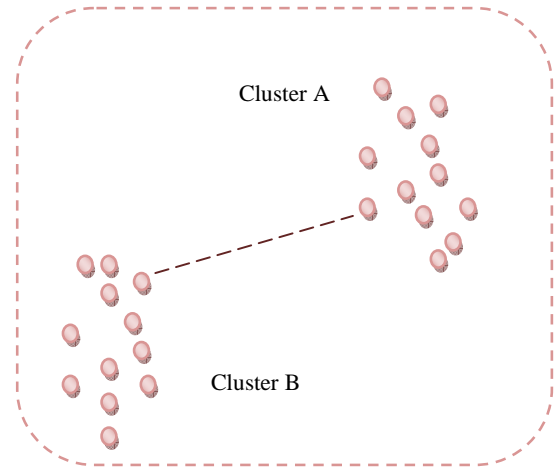

Fig. 1   Single linkage method.

## 3. Proposed method

In previous methods, field similarities were concerned directly and other operations were done on this obtained method, but in this method (Cluster Similarity Method) we will use cluster similarity of field in order to find the duplicate records. Initially the similarity of field is evaluated, and then clustering is done on them. In other words, each field is converted to some clusters, then similarity of records is calculated according to clusters of each group of fields and finally the last clustering is done in order to determine which records are placed in a cluster and which of them are similar.

3.1 Cluster similarity

Type errors and other cases result in duplicate records. Some of these defects are related to deficient entering of data, so that for example if one two-part word is in the form of AB and it is entered either A or B that their field similarity becomes like figure 2, in fact when we consider field similarity, AB has 0.5 similarity with A, but A does not have any similarity with B that is AB, when we consider cluster similarity, A and B are in one database, the fields of A and B are located in one cluster and in the next step, the more meaningful record similarity is gained and in fact this option decreases the rejection of error. In fact we fine some similarities by using present data in database that it seems that they do not have any similarity but they are the same.

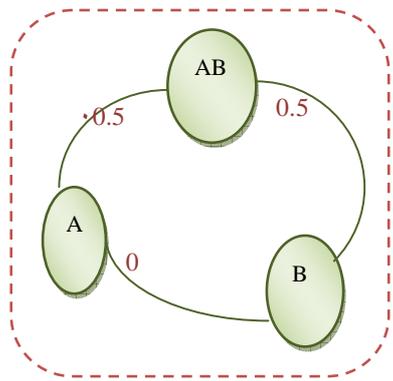

Fig. 2 Schematic presentation of field similarity.

In "Cluster Similarity Method", the cluster similarity is used in order to fine proper and meaningful similarity, so that it maximizes the accuracy and could find records which do not have more similarity and put them in one cluster. Clustering is done on each field, that each field is divided into a set of clusters. It may be the number of created clusters differs from one field into another field, for example that may become the name of 10 clusters and family name of 18 clusters. After finishing the clustering of fields, it may have the form of figure 3.

In figure 3 the parameters of X and Y is the representative of the present fields, example: X= name field and Y= last name field. Also x1, x2, x3 … related to X show correspondent records of that field, by assuming that x1 means the value that is in the field x and in record 1 and so on.

In "Cluster Similarity Method", we have three types of clustering: invalid cluster, empty cluster and valid clusters. If the present value in one field is invalid, it locates in invalid cluster. Invalid value means for example there is value of 125 in the name field that is invalid and or if there is the value of "ssss" in the field of national code that is invalid. If the field is empty and no value was entered, it locates in empty cluster. Other values that have proper format will locate in valid clusters that in this group the values are clustered based on their similarities.

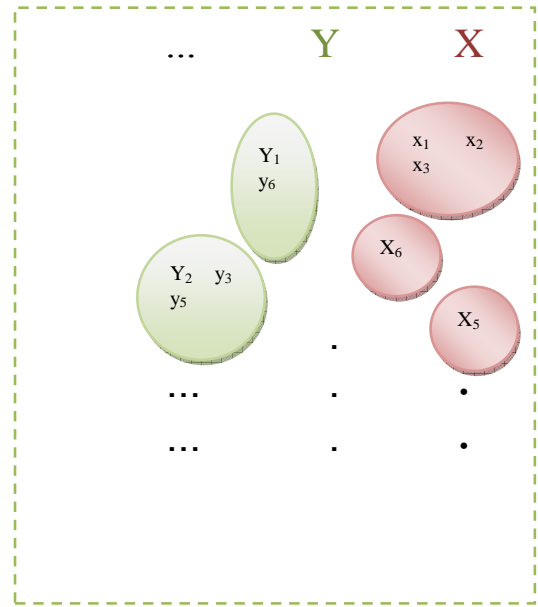

Fig.3 The way of fields' clustering.

## 3.2 Record similarity

After accomplishing the clustering on each field, in this step decision making is done according to cluster similarity about record similarity and the extent of similarity among records is specified.

The extent of record similarity is given by equation (4) that is a number between one and zero that one means completely same and zero means there is not any similarity.

$$Sr = \sum_{i=1}^{f} \frac{1}{K} Sc_i Df_i \qquad (4)$$

where Df shows the importance of fields that is calculated according to equation (7) and Sc show the degree of cluster similarity that is a number between one and zero that zero means these two clusters do not have any similarity and one means these two valued are in one cluster and also f means the number of present fields in a record. The value of K is determined according to the number of fields that are in invalid, empty and valid clusters and in other words, it is determined based on the validation of fields and it is calculated according to equation (5).

$$K = \sum_{i=1}^{f} k_i v_i \qquad (5)$$

f shows the number of present fields in one record and vi equals to the weight that is given to each field and the range of ki in this equation is between zero and one that is initialized according to the type of cluster by using equation (6).

$$k_i = \begin{cases} 0 & \text{Invalid} \\ 0 & \text{Empty} \\ 1 & \text{Valid} \end{cases} \qquad (6)$$

The second component of equation (4) is Df that shows the importance extent of fields that is given by equation (7)..

$$Df_i = k_i v_i \qquad (7)$$

That in this equation ki is obtained by equation (6) and vi equals to the weight that is given to each field. Since different fields have different values, we retain this balance by putting weight. For example in a database the value of family name may be more than address, so we give more weight to family name.

3.3 Record similarity

After finding the degree of similarity among records, clustering on records is done, so that similar records are located in one cluster. Selective clustering is a single linkage clustering that has better accuracy compared with other methods.

## 4. Conclusions

We introduce 3 criteria for evaluation that include R (Recall), P (Precision) and F1 (F-measure). Since final evaluation has been done on clustering, criteria also discuss both on the number of proper clusters and improper clusters, so that P and R are given by equations (9) and (10), respectively. And finally the value of F1 is calculated based on P and R like equation (10) [5].

$$R = \frac{\text{Number Of Common Clusters}}{\text{Number Of Manual Clusters}} \qquad (8)$$

The number of common clusters means how many clusters are there to make them same either manually or programmatic. The number manual clusters shows that these records were clustered in how many clusters manually.

$$P = \frac{\text{Number Of Common Clusters}}{\text{Number Of Program Clusters}} \qquad (9)$$

The number which is considered to the number of program clusters equals to the number of clusters that program form after final clustering.

$$F1 = \frac{2*P*R}{P+R} \qquad (10)$$

In hierarchical clustering, initially each data is put into one cluster and in each step the near clusters are combined in order to reach one unit cluster. Here we want to combine clusters to some extent that this amount shows that to somewhat similar clusters will be combined. In other words, it shows the maximum extent of similarity that clusters should have in order to combine. The proper selection of extent value has great influence on results, so one other parameter that is introduced here is Te and it shows the stop condition for single linkage clustering algorithm. The range of this value is between zero and one. We can obtain the proper value of Te for clustering by giving different values to Te and final testing of R, P and F1 that the results of these tests were shown in figures 4 and 5.

According to figure 4, as the value of Te increases up to 0.8, the value of F1 increases, so that in Te=0.8, the value of F1 reaches its highest value and when the value of Te becomes more that 0.8, the value of F1 decreases gradually. Therefore for "Cluster Similarity Method" Te=0.8 was set. Also according to figure 5, the value of Te in previous method in 0.8, obtained the highest value of F1, so for this method Te=0.8 was selected too.

We clustered database records in two ways, both with "Cluster Similarity Method" and with one of the previous methods which was described in section 2 and we presented the results in table 1. As table shows, the "Cluster Similarity Method" has better accuracy and F1 compared with previous method that approximately 24 percent improvement was reached.

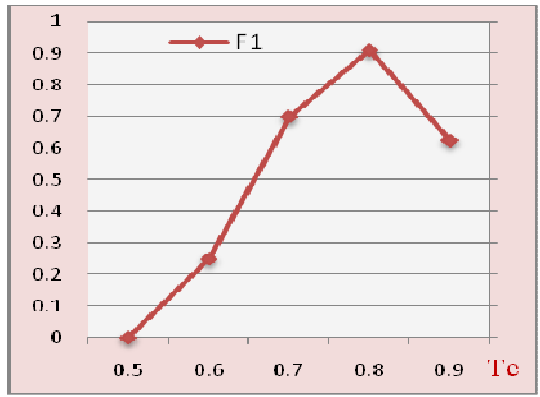

Fig. 4 Determination of stop condition's value (Te) for "Cluster Similarity Method".

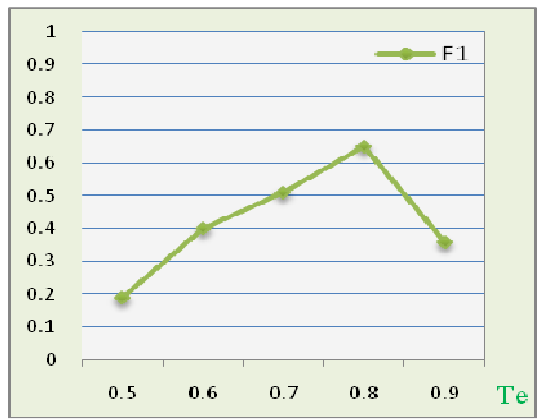

Fig. 5 Determination of stop condition's value (Te) for previous method.

Table 1: The results of methods' comparison

|  | *F1* | *P* | *R* |
|---|---|---|---|
| Cluster Similarity Method | 0.91 | 0.89 | 0.94 |
| Previous method | 0.65 | 0.62 | 0.67 |

## 5. Conclusions

In this method, cluster similarity was used; cluster similarity use present data in database in order to find similar fields and it is not based on direct similarity of fields. So it resulted in more logical clustering of similar records. Also the accuracy of calculation was increased, so that the accuracy of 80 to 95 percent was gained.

## References

[1] K. Elmagarmid, P. G. Ipeirotis and V. S. Verykios, "Duplicate record detection: A survey" IEEE Trans. on Knowledge and Data Engg., vol. 19, no. 1, pp. 1–16, 2007
[2] F. Maggi, "A Survey of Probabilistic Record atchingModels, Techniques and Tools", Scienti_c Report TR-2008.
[3] A. Furer, "Combining Runtime and Static Universe Type Inference" Master Project Report, Software Component Technology Group Department of Computer Science ETH Zurich, 2007.
[4] F. Mali, S. Mitra, "Clustering of symbolic data and its validation", Advances in Soft Computing, 2002.
[5] J. B. Santos, C. A. Heuser, V. P. Moreira and L. K. Wives, " Automatic threshold estimation for data matching applications", Elsevier, information sciences, 2010.

**Mohammad-Reza Feizi-Derakhshi:** was born in 1975. He received the B.Sc. degree in Computer Engineering from Isfahan University, Isfahan, Iran, in 1997, the M.Sc. degree in Computer Engineering from Iran University of Science and Technology, Tehran, Iran, in 2000, the Ph.D. degree in Computer Engineering from Iran University of Science and Technology, Tehran, Iran, in 2007. Since 2007, he is an assistant professor in the University of Tabriz, Tabriz, Iran.

**Aazde Rohany**: was born in 1980. She received the B.Sc. degree in Computer Engineering from Islamic Azad University, shabestar branch, Shabestar, Iran, in 2008, the M.Sc. degree in Computer Engineering from Islamic Azad University, shabestar branch, Shabestar, Iran, in 2011.